\documentclass{article}

\usepackage[final]{neurips_2019}

\usepackage[utf8]{inputenc} 
\usepackage[T1]{fontenc}    
\usepackage{hyperref}       
\usepackage{url}            
\usepackage{booktabs}       
\usepackage{amsfonts}       
\usepackage{nicefrac}       
\usepackage{microtype}      
\usepackage{graphicx}
\usepackage{amsmath}
\usepackage[table,xcdraw]{xcolor}
\usepackage{bbm}
\usepackage{natbib}

\title{Machine Learning for a Music Glove Instrument}

\author{%
  Joseph Bakarji\\
  Department of Energy Resources Engineering\\
  Stanford University\\
  \texttt{jbakarji@stanford.edu} \\
}

\begin{document}

\maketitle

\begin{abstract}
A music glove instrument equipped with force sensitive, flex and IMU sensors is trained on an electric piano to learn note sequences based on a time series of sensor inputs. Once trained, the glove is used on any surface to generate the sequence of notes most closely related to the hand motion. The data is collected manually by a performer wearing the glove and playing on an electric keyboard. The feature space is designed to account for the key hand motion, such as the thumb-under movement. Logistic regression along with bayesian belief networks are used learn the transition probabilities from one note to another. This work demonstrates a data-driven approach for digital musical instruments in general.
\end{abstract}

\section{Introduction}
\subsection{Motivation}

The possibilities in the design of electronic musical instruments are infinite. 
This is an unprecedented opportunity for engineering and musical creativity. 
However, music instrument designers have to experiment with different interfaces that they build and hardcode from scratch. 
This is a tedious process, and often leads to unintuitive instruments that have either too many or too little degrees of freedom. 
Therefore, optimizing the instruments design might take forever and never lead to any good music. 
This is what \cite{cook2001} called the ``curse of programmability'' in digital music instrument design.

To address this problem, a data-driven design approach of electronic musical instruments is proposed. 
In particular, an electronic glove with sensors is trained on a digital piano by a performer wearing the glove while playing on the keyboard. 
Once trained, the glove is used on any surface to play music, mapping the sensors of the glove to full-keyboard midi messages. 
This is a novel approach to instrument design that has never been used before, to the best of my knowledge.

\subsection{Literature Review}
This being said, related work has been done in the past.
\cite{voutsinas2017mi} and \cite{erad2015glovesense} used a glove with sensors to play music with various degrees of complexity. 
Their gesture-to-note mappings were mostly hard coded.
With the recent advent of virtual reality, hand gesture interface devices have been developed by \cite{zimmerman1987hand} and \cite{kim20093} for game-play and gesture recognition with some form of machine learning.
Recently, there has been increasing interest in combining machine learning and music in an interactive setup as demonstrated by \cite{fiebrink2011real} and \cite{fiebrink2009meta}.
However, there is still much to be done beyond proof of concepts for interactive musical instrument design with machine learning.

Imitation learning as presented by \cite{schaal2003computational} provides a perfect framework for training newly designed instruments with other electronic instruments like a keyboard.
Imitation learning in this context takes the sensor inputs from the glove as observations, and outputs midi notes as actions.

\section{Problem Setup}

\subsection{Task Definition}
A mapping between the glove sensors (used as inputs) and the piano keyboard (used as true lables) is sought. Specifically, we seek to train a glove that has 5 pressure sensors, 5 flex sensors, and an IMU to mimic a piano performance using 88 keyboard keys. This is in principle an ill-defined problem in which only 5 fingers are mapped to 88 keys without explicit knowledge of the hand position. Only a relative position can be estimated by the motion sensor, and thus the first reference note is assumed. Once the glove is trained by a piano performer, the glove can be used on any table as if playing a virtual keyboard. The task is to leverage both data and AI algorithms to accomplish this.

\begin{figure}[htbp]
\begin{center}
\includegraphics[width=12cm]{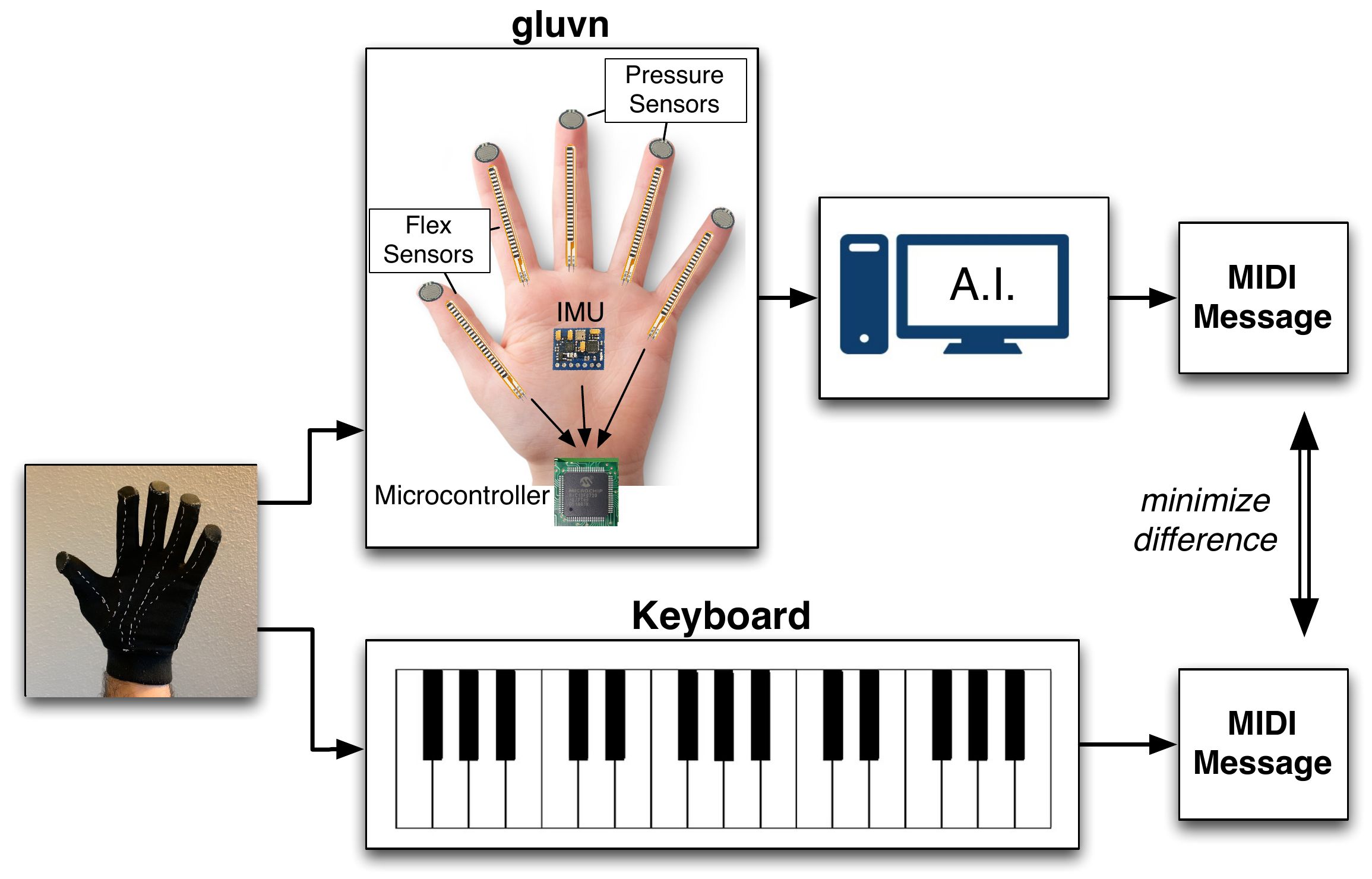}
\caption{{\small Illustration of glove learning to play music from keyboard}}
\label{gluvill}
\end{center}
\end{figure}

Teaching a glove to generate notes has many levels. First, the glove has to be either trained or hardcoded to trigger notes when the pressure applied on the finger reaches a certain threshold. Once the threshold is reached, the volume of the note has to be specified to be sent as part of the midi message. Finally, the next note has to be chosen based on the present note and possible previous notes. In this project, it is assumed that only one note is triggered at a time, and that we're only seeking a melody. This can of course be generalized using more sophisticated learning methods and musical constraints.

\subsection{Inputs and Outputs}
Formally, let the pressure, flex and IMU readings be defined by $S^p_{1..5}(t_i)$, $S^f_{1..5}(t_i)$ and $S^a_{1..6}(t_i)$ (where the subscripts indicate 5 flex and pressure readings, and 6 IMU, i.e. gyroscope and accelerometer readings) respectively, given as a function of a discrete time $t_i$ that depends on the sampling rate, where $i$ is defined to be relative to the present moment (i.e. $i<0$ for past readings and $i=0$ for the present reading). A sample of 5 pressure readings and 1 flex sensor reading is shown in Fig.~\ref{sample}.

The inputs to the algorithm (or feature space) are a subset of the sensor readings, along with a subset of the previous notes already played. In general, these can be summarized by the input set given by 
\begin{align*}
\mathcal I^{\{pfa\}}_{nm}&= [S^{p(n)}, S^{f(n)}, S^{a(n)}, \hat N^{(m)}, F^{(m)}]\\
 S^{\alpha(n)} &= [S^\alpha(t_{-n}), S^\alpha(t_{-n+1}), \dots, S^\alpha(t_0)] \\
\hat N^{(m)} &= [\hat N_{-m}, \hat N_{-m+1}, \dots, \hat N_{-1}] \\
F^{(m)} &= [F_{-m}, F_{-m+1}, \dots, F_0]
\end{align*}
where $\hat N_{-i} = (\hat \eta, \hat v)_{-i} \in [0, 127]^2$ is the $i$-th previous note midi number, $\hat \eta_i$, played by the algorithm along with its velocity, $\hat v_{-i}$, and $F_{-i}$ the corresponding previous $i$-th finger. 
The piano keyboard is considered the real world or the environment outputting a sequence of ground truth notes $N_{-i} = (\eta, v)_{-i}$.

Given the variables above, once a finger is triggered, an optimal mapping $h_\theta(\cdot)$ is sought to output the next note based on the previous $n$ sensor readings and $m$ notes 
\begin{equation}
\hat N_0 = (\hat \eta_0, \hat v_0) = h_\theta(\mathcal I^{\{pfa\}}_{nm})
\end{equation}
This is done by minimizing an objective function that is a function of the distance between the predicted note and the actual note $||h_\theta(\mathcal I^{\{pfa\}}_{nm}) - N_0||$.
In the context of imitation learning, that optimal is simply found by a probability transition matrix and the action (or note) is sampled from a probability distribution learnt from the training set.

\subsection{Infrastructure}
\subsubsection{Data Collection}
The data is collected by connecting a midi keyboard and the glove micro-controller to the same computer. A python program reads both serial ports simultaneously using multi-threading and saves both sensors and keyboard input data. The baselines are a set of simple scales and 5-finger-to-5-notes performances. As the algorithm becomes more powerful, the accuracy of increasingly more complex performances is expected to become $\ge 90\%$.

\begin{figure}[htbp]
\begin{center}
\includegraphics[width=12cm]{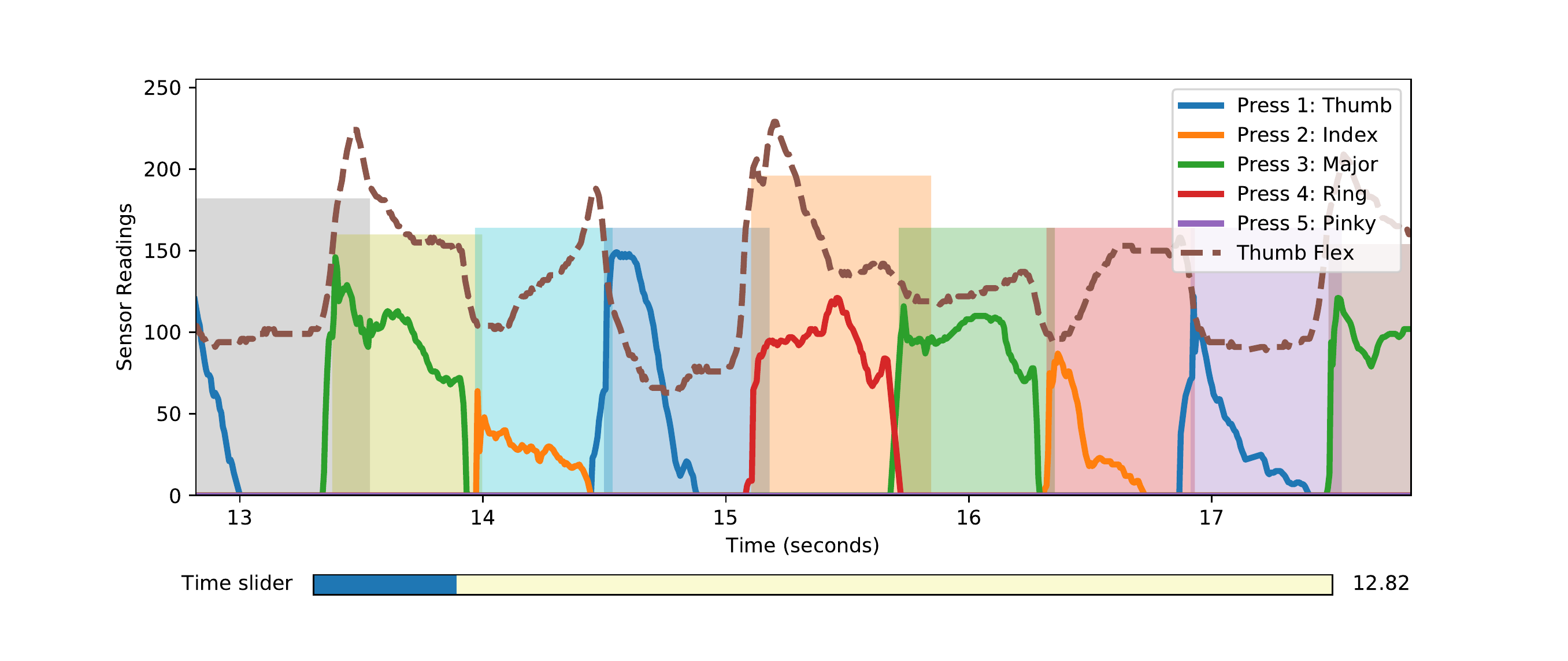}
\caption{Sample of 5 Pressure and 1 flex sensor readings, with the corresponding volume of notes from the keyboard}
\label{sample}
\end{center}
\end{figure}

There are some issues when it comes to data collection. The sampling rate is around 150 Hz but there are packet losses at the serial port delaying the readings up to 0.2 seconds. Given the irregularity in the sensor readings, time has to be taken into account when learning outputs based on vector inputs. However, this is not taken into account.
Furthermore, the sensor and keyboard readings are assumed to be perfectly synchronized but the data shown in Fig.~\ref{sample} sometimes proves a lag between the two readings.

The flex sensors have also been observed to drift with time making the training set not homogeneous. With time, flex sensors become bent permanently, making the mapping between sensor and the actual state of the hand motion not unique. This is fixed by shuffling the recorded data, and recalibrating the sensors at every run. However, the recalibration is not perfect which explains some misclassification in the learning tasks later.

The finger input corresponding to each label (note) is determined based on the maximum integral of the pressure sensors.
That is, the pressure sensor that has the highest integral from trigger to release is the one playing the note
\begin{equation*}
F = \text{argmax}_{i}\int_{\text{trigger}}^{\text{release}} S^p_i(t) dt
\end{equation*}
where $S^p_i(t)$ is the pressure reading of finger $i$.

The data is also processed to learn a smaller problem. First, all training sets start with the note C, and are played on the C scale (white notes). To facilitate the analysis and learning, the C-major midi notes are transformed to index numbers ranging from 25 to 90. This is akin to just numbering the white notes on a keyboard.

\subsubsection{Generalization}
Another issue is generalization. The purpose of this study is to first learn the notes on the keyboard, then use the glove on a table. This assumes that the movement of the hand on the keyboard will be the same as the one on the table. But that's not the case in reality. In particular, when the glove triggers a keyboard note, there is a spike in the pressure sensors when the glove first touches the note, and then a further increase when the note is fully pressed. A different pressure dynamic happens when the glove is played on a hard surface. This means that learning the mapping between sensors and midi output on the keyboard will not generalize to playing on a hard surface. This makes learning note triggers hard or impossible without a model of note dynamics. Learning the velocity of the note given this pre-trigger spike artifact would also be tricky. 

This is why the focus in this study is to learn note intervals, given a hard coded way of triggering notes and computing their velocity. In addition, the testing will be performed on the data collected from playing on the keyboard rather than tested on a hard surface, as explain in section \ref{results}.

\section{Modeling Approach}

\subsection{HMM}
Given a sequence of fingers $F_i$ and notes $N_i$, the aim is to maximize the conditional probability $\mathbb P(N_i | N_{i-1}, F_i)$, or the probability of having note $N_i$, given the previous note and the current observed finger playing. This is a hidden Markov model (HMM), with $F_i$ as observations, and $N_i$ as hidden states. In this case, the domain of $F_i$ is $[1..5]$ and that of $N_i$ is, in general, $[1..88]$. Once the transition probabilities $\mathbb P(N_i | F_i)$ and $\mathbb P(N_i | N_{i-1})$ are learnt from training data, one can infer the new state $N_i$ at every given ``measurement'' $F_i$. 

There are 2 potential issues with this model. First, the dynamics of playing the piano $\mathbb P(N_i | N_{i-1})$ will be arbitrarily dependent on the style of the glove user, and might not generalize to a meaningful model. Second, the transition matrix representing $\mathbb P(N_i | F_i)$ is of size $88 \times 5 = 440$, and that of $\mathbb P(N_i | N_{i-1})$ of size $88 \times 88 = 7744$, making the inference problem computationally expensive, and the training unlikely to generalize. 

\subsection{A Hand Plays Intervals}
To reduce the problem complexity, it is useful to note that when playing the piano the movement of the hand is invariant to translation. Thus the sensor readings only reveal the intervals played $\Delta N_i = N_i - N_{i-1}$, rather than the specific notes $N_i$. Redefining the states accordingly, and including the previous finger, the probability we seek is $\mathbb P(\Delta N_i = N_i - N_{i-1} | F_{i-1}, F_i)$. The new transition matrix has a theoretical size of $5 \times 5 \times 88 \times 2 = 4400$ but in practice the matrix is much smaller because very large hand jumps in piano are rare compared to the adjacent ones. An even simpler model is tested for comparison, involving the probability $\mathbb P(\Delta N_i | \Delta F_i)$, assuming that the difference in notes only depends on the difference in fingers. In this setup, the input is given by $\mathcal I_{1}$ from the above definition.

\begin{figure}[htbp]
\begin{center}
\includegraphics[width=12cm]{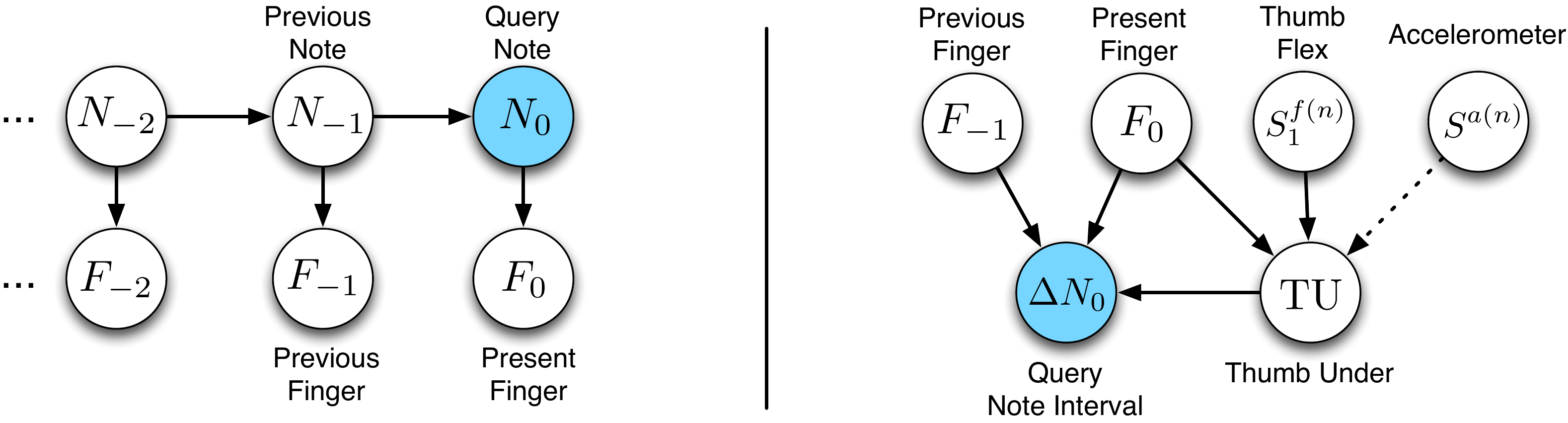}
\caption{Bayesian network models for problem of predicting next note $N_0$ or note interval $\Delta N_0$. A typical HMM model on the left. Model used in this problem on the right.}
\label{bns}
\end{center}
\end{figure}

\subsection{Accounting for Thumb Unders}
Given only fingers playing as input features, the algorithm has no way of knowing when a ``thumb under'' event occurs (when the thumb is passed under the other fingers to go up or down the scale). This can be accounted for by adding the flexing of the thumb $S_1^{f(n)}$ in the feature space, with $n$ being a hyperparameter to be optimized (turns out to be around 10). 

To do so, the task of predicting $\Delta N_0$ is split in two as shown in Fig.~\ref{bns}. First, we learn the predictor $\mathbb P(TU | F_0, S_1^{f(n)})$ using logistic regression to classify weather a thumb under occurred or not. And then we predict $\Delta N_0$ by adding the binary feature $TU$ to the feature space to learn $\mathbb P(\Delta N_0 | F_{-1}, F_0, TU)$. This will only double the size of the matrix while adding a lot of information about the movement of the hand. 

The feature space $\phi(I^{\{pfa\}}_{nm}) = S_1^{f(n)}$ in the $TU$-classification problem is of size $n$ for a given present finger $F_0$, and the corresponding predictor is a binary variable $TU$ that takes the value
\begin{equation}
\hat {TU} = \mathbbm{1}[\sigma(S_1^{f(n)} \cdot W_{F_0}) \ge 0.5]
\end{equation}
This means that in practice we have 5 binary classifiers $\hat{TU}$, each with weights $W_{F_0}$, one for each present finger $F_0 \in \{1 .. 5\}$.

The thumb-under true labels are determined according to the known values of present and previous fingers and notes using the following relation
\begin{equation*}
TU = \mathbbm{1}[ (F_0 - F_{-1}) (N_0 - N_{-1}) < 0]
\end{equation*}
That is, if fingers and notes are changing in opposite directions, fingers should be crossing (a thumb under). Note that the data is generated with this in mind.

\section{Results and Error Analysis}\label{results}

The first 80\% of each run is taken as a training set and the last 20\% as part of the dev set. All training sets are combined but tested on each file separately to monitor the accuracy as a function of the (piano) performance complexity. As shown in table \ref{acctable}, the complexity of performance increases from the use of 5 notes only (CDEFG) to an improvisation with predictable intervals (not more than 5 notes). The scales training set is a simple scale on C major, the Bach menuet is a short piece, the Improv-pred-int set is a random improvisation on C major, and the Improv-nonPred-Int set is the most general improvisation performances including all possible intervals. Note the accuracy is much lower on this last data set given both the lack of enough training examples and the large bias due to the lack of enough features that describe hand movement. Including the accelerometer data is expected to improve the performance.

The increase in accuracy can be see to be a function of the number of features; more features is better, especially when accounting for the presence of a thumb-under movement. The interval of fingers also seems to be better predicted when the fingers are known rather than the finger interval alone. This is particularly true for thumb-under movements which happen on fingers (3, 1) with note interval 1, but never on fingers (4, 2) which most gives a note interval of -2.

\begin{figure}[htbp] 
\begin{center}
\includegraphics[width=6.5cm]{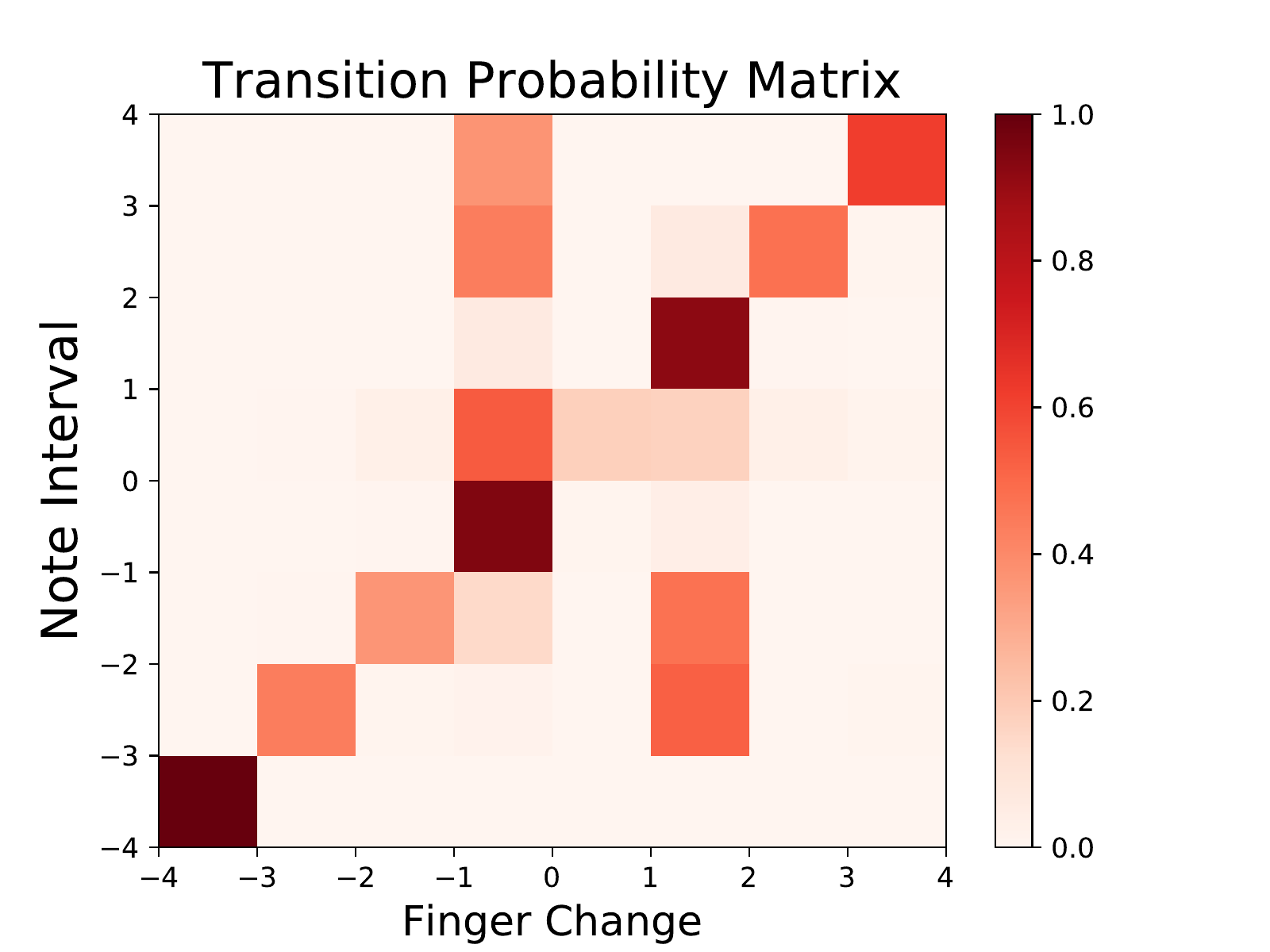}
\includegraphics[width=6.5cm]{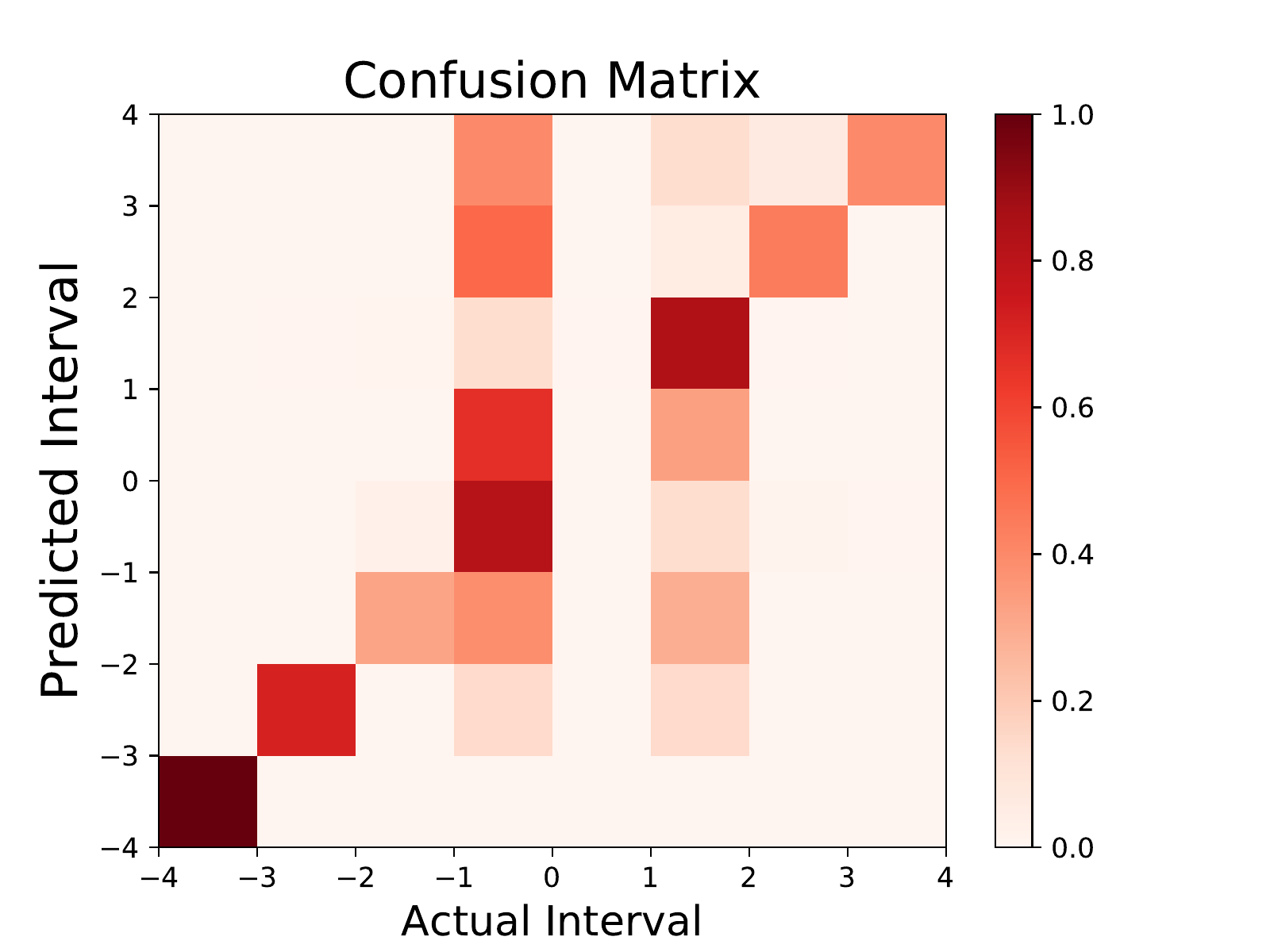}
\caption{Left: transition matrix for the probability $\mathbb P(\Delta N_0 | \Delta F_0)$ and a training set with predictable intervals (i.e. not greater than 5). Left: Confusion matrix (normalized row-wise) for feature space $(F_0, F_1, TU)$.}
\label{probas}
\end{center}
\end{figure}

The transition probability for the simplest feature space is shown in Fig.~\ref{probas} on the left. The results show a concentration around the diagonal which means that the most frequent correlation between fingers and notes is a one-to-one mapping. On the right, Fig.~\ref{probas} shows the normalized confusion matrix with the feature space $(F_0, F_1, TU)$, also having concentrated predictions along the diagonal now indicating the correct predictions. It can be seen that there is an improvement in predicting the thumb under for interval $(1, 3) \rightarrow -1$. However, the performances seem to lack enough $(0, 0)$ (no note change) occurrences. Many of those errors are due to noisy sensors requiring more filtering and data cleaning in the future.

\begin{figure}[htbp] 
\begin{center}
\includegraphics[width=6.5cm]{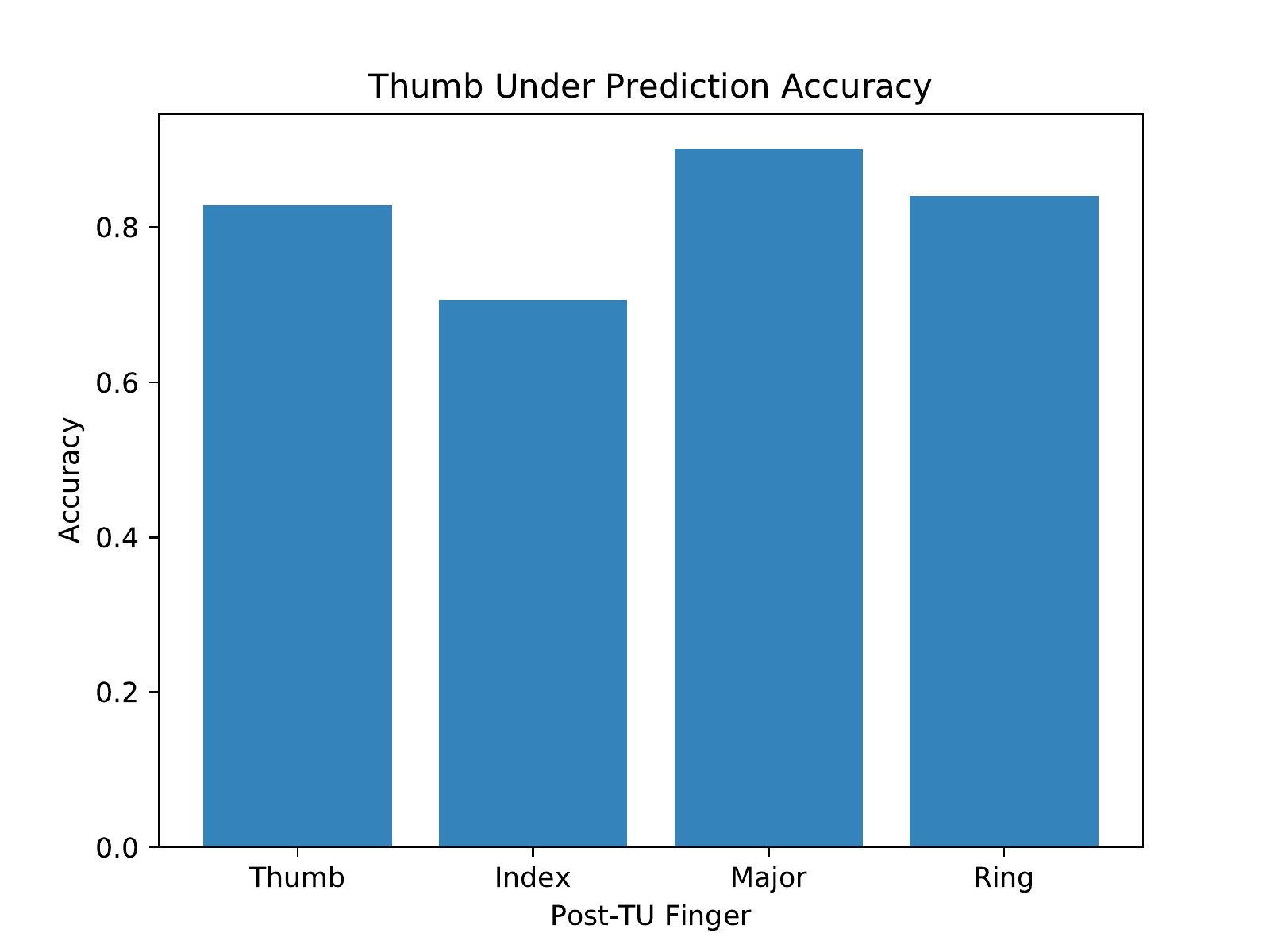}
\includegraphics[width=6.5cm]{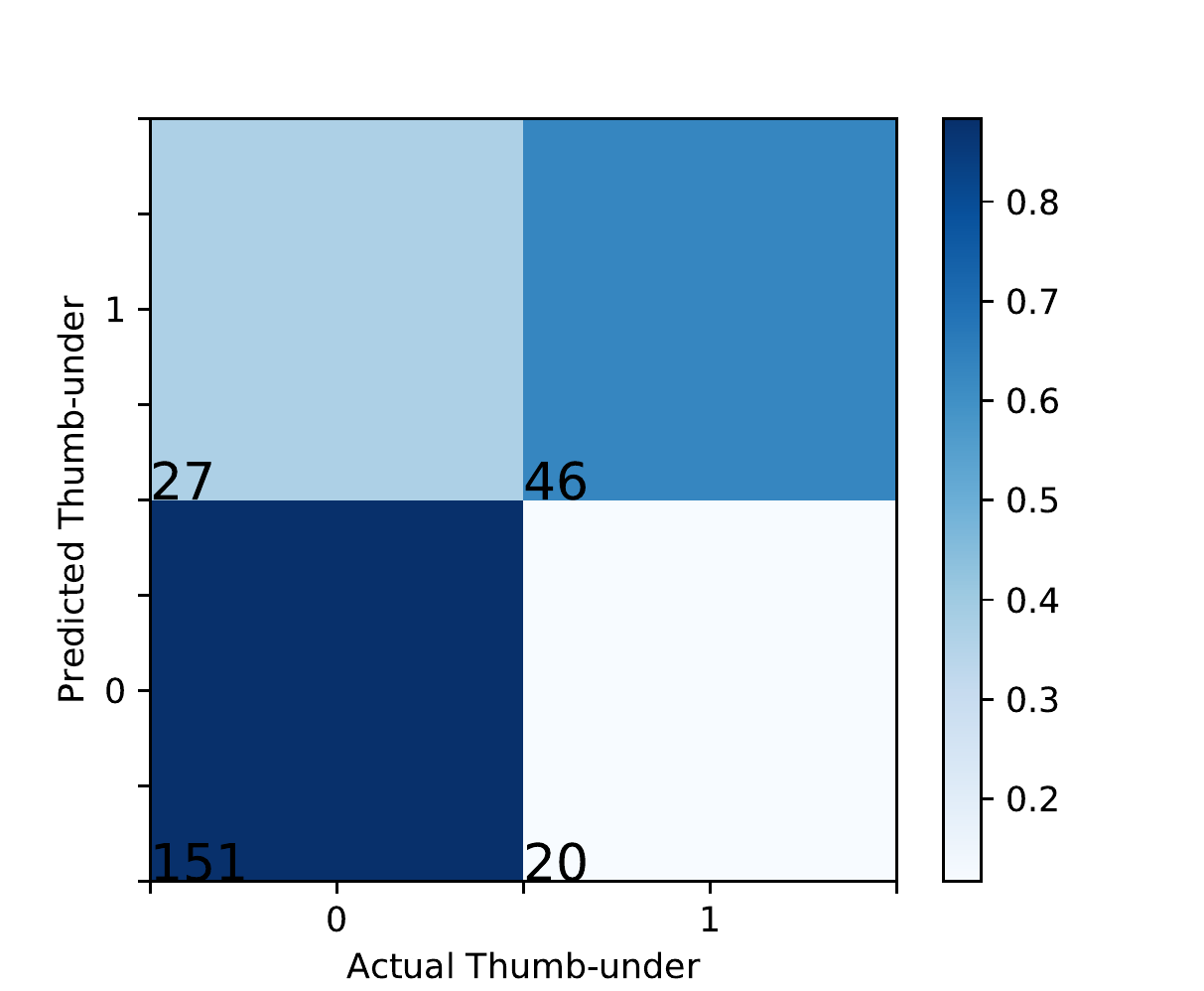}
\caption{Left: Accuracy of thumb under movement prediction given the present finger $F_0$. The pinky finger has been omitted because it never happens after a thumb under in practice. Right: confusion matrix for the thumb as present finger, showing around 50\% of false positives suggesting a lack of positives in the training set.}
\label{tupred}
\end{center}
\end{figure}

Figure \ref{tupred} shows the accuracy of thumb-under predictions. Using logistic regression, a thumb-under event is predicted up to 90\% of the time on average. However, as shown in the confusion matrix, the number of negatives is much larger than the positives, making most correct predictions true negatives. Improving upon these predictions, more data has to be collected in the future.

\begin{table}[h]
\label{acctable} 
\begin{center}
\begin{tabular}{c|c|c|c|}
\cline{2-4}
\multicolumn{1}{l|}{} & \multicolumn{3}{c|}{\textbf{Feature Space}} \\ \hline
\multicolumn{1}{|c|}{\textbf{\begin{tabular}[c]{@{}c@{}}Performance\\ Complexity\end{tabular}}} & \multicolumn{1}{l|}{$(F_0 - F_1)$} & \multicolumn{1}{l|}{$(F_0, F_1)$} & \multicolumn{1}{l|}{$(F_0, F_1, TU)$} \\ \hline
\multicolumn{1}{|c|}{CDEFG}                & 0.92 & 0.96 & 0.96 \\ \hline
\multicolumn{1}{|c|}{CDEFG-Rand}      & 0.81 & 0.84 & 0.84\\ \hline
\multicolumn{1}{|c|}{Scales}                 & 0.70 & 0.80 & 0.85 \\ \hline
\multicolumn{1}{|c|}{Scales-Random}  & 0.69 & 0.75 & 0.82 \\ \hline
\multicolumn{1}{|c|}{Bach Menuet}      & 0.70 & 0.72 & 0.80 \\ \hline
\multicolumn{1}{|c|}{Improv-Pred-Int}  & 0.55 & 0.62 & 0.70 \\ \hline
\multicolumn{1}{|c|}{Improv-nonPred-Int}  & 0.40 & 0.50 & 0.56 \\ \hline
\rowcolor[HTML]{FFCCC9} 
\multicolumn{1}{|c|}{\cellcolor[HTML]{FFCCC9}{\color[HTML]{333333} Avg Accuracy}} & {\color[HTML]{333333} 0.69} & {\color[HTML]{333333} 0.76} & {\color[HTML]{333333} 0.80 } \\ \hline
\end{tabular}
\caption{Table of accuracy with increasing performance complexity and increasing feature space complexity. (The average accuracy across performances is not weighted by the length of performance)}
\end{center}
\end{table}

The results in table~\ref{acctable} show that a more complex feature space improves accuracy. However non-predictable intervals do very poorly with only 5 pressure sensors and 1 flex sensor. Improving upon those results will require including the full set of sensors and a more sophisticated learning algorithm.

\section{Conclusion and Future Work}
A bottom-up learning approach for mapping hand motion to midi output has shown relative success.
The challenges of prediction speed, limited data, and feature selection have been addressed and the corresponding error analyzed.
The study introduces a novel and general framework for designing an electronic instrument by imitating the movement on another if the designer has a way to use both instruments simultaneously.
This framework can be used as an intermediate step in a more complex design.

A limitation in this study is that designing the feature space and learning some of the features before prediction is tedious task.
A more general and powerful approach would be to use a deep learning predictor in which features are learned automatically. 
But this requires much more training data which wasn't available in this study.
Once more data is collected, a simple possibility is to use a CNN followed a fully connected network. The first part would take the last $n$ flex and accelerometer data to encoding them in $m$ output features. Then these features are used along with the fingers and previous notes as inputs to a fully connected network to predict the interval $\Delta N_0$.
Another possibility is to simply use an RNN network given the sequential nature of the data.

Perfecting the design of the instrument involves adding constraints such as scale, chord and interval ranges which can be done with many AI techniques such as factor graphs and bayesian networks.
Future work will include two gloves, a pedal, all possible notes (with scales as a learned feature) with the goal of having a full ``virtual keyboard'' that generates notes according to the movement of the virtual pianist.

\bibliography{draft}
\bibliographystyle{plainnat}

\end{document}